\documentstyle[psfig,12pt]{article}

\begin{document}

%\maketitle

\begin{center}

\Large
{\bf A Spectral Diagnostic for Density-Bounded H~{\sc ii} Regions}

\normalsize
J. Iglesias-P\'{a}ramo$^{1}$ and C. Mu\~{n}oz-Tu\~{n}\'{o}n$^{2}$ 

$^{1}$Laboratoire d'Astrophysique de Marseille,
Traverse du Siphon - Les Trois Lucs, 13376 Marseille, France

$^{2}$Instituto de Astrof\'{\i}sica de Canarias,
E-38205 La Laguna, Tenerife, Spain 

email: jorge@astrsp-mrs.fr, cmt@ll.iac.es

\end{center}

%\doublespace

%\normalsize

\section{Abstract}
The existence of density-bounded H~{\sc ii} regions in spiral galaxies is
supported by means of a spectral indicator based on the intensity of the [O~{\sc
i}] $\lambda$6300 \AA\ forbidden line.
A grid of photoionization models providing spectral information of
density-bounded nebulae is presented in order to test the validity of our
indicator. 
%A grid of models for ionized nebulae that
%provide a spectral diagnostic for the detection of density-bounded H{\sc
%ii} regions in galaxies. We test the validity of an indicator of
%density-bounding based on the intensity of the [O~{\sc
%i}] $\lambda$6300 \AA\ forbidden line and stress the importance of having good
%measurements or reliable upper limits to the intensity of this line in order
%to be able to discriminate between density- or ionization-bounded 
%H~{\sc ii} regions. 
The indicator is applicable to typical observed
H~{\sc ii} regions with solar or higher metallicities, and for the range of
ages  observed in normal H~{\sc ii} regions. 
When applying the diagnostic to a large sample of
H~{\sc ii} regions of spiral galaxies 
taken from the literature, we find that a fraction of the selected H~{\sc ii}
regions with  emission detected in the [O~{\sc i}] $\lambda$6300 \AA\ line
turn out to be consistent with the predictions of density-bounded
photoionization models. Preliminary estimates of the fraction
of Lyman continuum photons escaping ranges from 20\% to 40\%. The
number of density-bounded H~{\sc ii} regions could be even larger after
confirmation of those regions for which the [O~{\sc i}] line was not detected,
provided they are good density-bounded candidates. We stress the
importance of having combined information on both good measurements or
reliable upper limits for the intensity of this line, and good determinations of
the H$\alpha$ luminosity for complete samples of H~{\sc ii} regions in galaxies
in order to make proper estimations as to whether UV photons escaping from H~{\sc
ii} regions are the main source of ionization of the diffuse ionized medium in
galaxies.

\section{Introduction\label{intro}}

H~{\sc ii} regions are clouds of ionized gas surrounding young
massive stars. Their spectra are dominated by emission lines
corresponding to H, He, and some forbidden lines of heavier elements. Their
physics is well known, and has been reviewed by several
authors (e.g., Spitzer 1978; Osterbrock 1989). In the classical view,
 H~{\sc ii}
regions are spheres of ionized gas illuminated by a star or a
stellar cluster emitting Lyman continuum photons. The extent of
the region is determined by the ionization front, where there is a balance 
between the number 
of photoionizations and recombinations of hydrogen atoms. 
When the H~{\sc i} cloud is large enough to consume all ionizing photons, the
ionization front is trapped, and the 
H~{\sc ii} region is said to be ionization bounded. Otherwise, 
there is no ionization front and  some Lyman continuum photons eventually
 leak from the parent cloud. The region is then considered to be
 density bounded.

The real picture shows that stars are normally born in the inner cores of
 dense molecular clouds. Thus, density gradients
are expected along the paths of Lyman continuum photons. The steeper the
gradient, the larger the external radius of the ionized region is. During the
 expansion phase, for a density gradient larger than about $R^{-1.5}$,
 the ionization front is not trapped and many Lyman continuum photons
 are expected to leak from the parent cloud (Franco, Tenorio-Tagle, \&
Bodenheimer 1990).

From optical emission-line and
pulsar dispersion measurements in the Milky Way, we
know now that most of the ionized gas is located
in very low density (more than ten times lower than in classical H~{\sc ii}
regions) clouds far from star clusters, even several hundreds of
parsecs above galactic disks. This ionized component would be
mixed with the neutral one and would constitute one-third of the total
mass of the interstellar medium. So, although
recombinations are mainly associated with star clusters, most of the mass
of H~{\sc
ii} regions is associated with an extended low density component (see
Reynolds 1993 for a review of the main properties of the diffuse
interstellar medium in our Galaxy).
Observations of diffuse ionized gas in the disks of spiral galaxies has
 led to the conclusion that many ionizing photons should escape from
 the H~{\sc ii} regions, producing a
diffuse ionized component (Ferguson et al. 1996).

Several authors have reported the existence of such density-bounded
H~{\sc ii} regions. 
The observational evidence is based on different considerations such as the 
non-correlation between the stellar and H$\alpha$ luminosities measured for
H~{\sc ii} regions (Oey \& Kennicutt 1998), the change in the slope of
the H$\alpha$ luminosity function for the brightest H~{\sc ii} regions
(Beckman, Rozas, \& Knapen 1998), changes in the $L$(H$\alpha$) vs.
$\sigma$ diagram (Fuentes-Masip et al. 2000) for 
H~{\sc ii} regions in irregular galaxies, or discordant emission lines with
respect to photoionization models (Castellanos et al. 2002). 
%However, no attempt has been made so far 
%to address this issue using the spectra of H~{\sc ii} regions.

We address the topic by studying the very
faint [O~{\sc i}] $\lambda$6300 \AA\ emission line. Since most of the [O~{\sc
i}] emission
arises from collisional excitations by thermal electrons
via the charge exchange reaction $\rm{H}^{+} + \rm{O}^{0}
\longleftrightarrow \rm{H}^{0} + \rm{O}^{+}$, 
its intensity is a measure of the neutral hydrogen content within the
ionized regions.
Although  in typical H~{\sc ii} region conditions the charge exchange has a rate
comparable with recombination in converting O$^{+}$ to O$^{0}$, charge exchange dominates at the outer edges of the nebulae because of the higher density of
H$^{0}$ (see Osterbrock 1989).
Therefore, the detection of the [O~{\sc i}] line in the
spectrum of an H~{\sc ii} region indicates that most
(if not all) of the Lyman continuum photons are absorbed by the nebula. In an
ideal H~{\sc ii}--H~{\sc i} discontinuity there would be a shell of [O~{\sc
i}] delimiting the border of the H~{\sc ii} region.

In this paper we describe a method for determining whether H~{\sc ii} regions with
solar or higher metallicity are density or ionization bounded, based on the
intensity of the [O~{\sc i}] $\lambda$6300 \AA\ line.
A grid of photoionization models has been constructed and
is described in Section \ref{model}.
Section \ref{didi}
shows diagnostic diagrams applied to the output spectra in
order to explore the influence of escaping Lyman continuum photons in the
spectra of H~{\sc ii} regions. 
Finally, Sect. \ref{discon} contains some discussion of the main
results of the paper and their application to a sample of H~{\sc ii}
regions in nearby spiral galaxies.

\section{The Grid of Models \label{model}}

A grid of models was constructed using the latest version of the
photoionization code CLOUDY (see Ferland 1993 for details). Each
 model is parameterized by the hydrogen density, the ionization parameter, the shape of the
central stellar continuum---including age and the slope of the initial mass
function (IMF) effects---and the chemical composition. Table~\ref{models}
shows the different values adopted.

Given the lack of geometrical information available for most H~{\sc ii}
regions in spiral galaxies, and that there exists a relationship
between the number of ionizing photons, the density of the nebula and
the filling factor\footnote{$u \propto
(Q_{\rm{H}}n\epsilon^{2})^{1/3}$, where $u$ is the ionization
parameter, $n_{\rm{H}}$ is the hydrogen density, $\epsilon$ is the
filling factor and $Q_{\rm{H}}$ is the number of ionizing photons.},
fixed values of the hydrogen density and the
ionization parameter were assigned to each model. 
The number of ionizing photons is a free parameter, as are some geometrical quantities.
The hydrogen density was set at 10 cm$^{-3}$ for all models below
the limit for collisional de-excitation of the important cooling
lines, and the ionization parameter ($\log u$) ranged from $-2$ to $-4$
(inclusive).

The chemical composition was chosen to cover a wide range of oxygen 
abundances, from 0.1\ $Z_{\odot}$ to 2\ $Z_{\odot}$,
according to the solar abundance of oxygen relative to
hydrogen defined by Anders \& Grevesse (1989). The relative
abundances of the other metals were scaled to that of  oxygen 
following the prescription for H~{\sc ii} regions (see Ferland 1993, for
details). The effects of dust were not considered.

Finally, we chose the spectral energy distributions (SEDs) from
Leitherer \& Heckman (1995) as ionizing sources, in such a way that
their metallicities matched those of the ionized gas. The
IMFs chosen were those of Salpeter ($\alpha = -2.35$), with upper limits of 30 and
100~$M_{\odot}$, and  Miller Scalo ($\alpha = -2.35$), with $M_{\rm up} = 100~M_{\odot}$.
Given that the shape of the ionizing continuum changes
with the age of the cluster, we run the code for ages between 1 and
10~Myr, with intervals of 1~Myr.

%The grid of models chosen covers the whole parameter space
%of the observed H~{\sc ii} regions.
%We classify two cases: ionization-bounded and density-bounded models. For
%the first case, CLOUDY considers all layers needed to account for all the
%Lyman continuum photons produced by the ionizing source. The
%density-bounding effect is reproduced by removing the most external ionized
%layers. The number of photons lost depends on the number of layers
%neglected. In our case, for density-bounded models, external layers
%are removed until the neutral oxygen abundances become an order of
%magnitude lower than in ionization-bounded models for the same
%input parameters.

%*******************************************

%Aqui explicar como hacemos los modelos density-bounded y como podemos calcular
%la fraccion de fotones que escapan...

The grid of models chosen covers the whole parameter space of the observed
H~{\sc ii} regions. Given that we want to study the effect of density-boundedness
in the spectra of H\ {\sc ii} regions, we produced both ionization-bounded and
density-bounded models. The ionization-bounded models are those for which the
radiation transfer is computed until the temperature of the nebula is 2500~K. At
this point the fraction of ionized hydrogen is negligible and we may consider
all the Lyman continuum photons to have been absorbed by the nebula. After
that, for each ionization-bounded model we produce a density-bounded model with
the same input parameters as the corresponding ionization-bounded model, but we
stop the computation of the radiative transfer at the point where the abundance
of O$^{0}$ reaches one-tenth of the value measured at the border of the
corresponding ionization-bounded model. Of course, the fraction of escaping
photons is not the same for all models and strongly depends on the value of $u$
assumed for each model. At the point we stop the computation of the radiative
transfer across the nebula, CLOUDY provides all the required information to
estimate the fraction of escaping photons as well as the emergent spectrum for
each density-bounded model.

%*******************************************

\section{Diagnostic Diagrams\label{didi}}

First, we tested the reliability of the models with a
classical diagnostic diagram for observed H~{\sc ii}
regions. Figure~\ref{o2o3} shows the diagram representing $\log$ 
([O~{\sc ii}] $+$ [O~{\sc iii}])/H$\beta$~{vs.}~[O~{\sc iii}]/[O~{\sc
ii}]. The shaded region corresponds to the locus occupied by the
H~{\sc ii} regions observed  by McCall, Rybski, \& Shields (1985).
The open and close squares correspond to ionization-bounded
and density-bounded models, respectively, chosen arbitrarily from our set. 
It is clear from the figure that both ionization- and density-bounded models 
coexist within the region occupied by the observed H~{\sc ii}
regions. This means that the strongest emission lines of oxygen do not provide
enough information to discriminate between density-bounded and
ionization-bounded H~{\sc ii} regions.

With the aim of finding an indicator of photon leakage in H~{\sc ii} regions,
as mentioned in Sect.~1, we focus our work on the
forbidden  [O~{\sc i}] $\lambda$6300\AA\ emission line. In 
Fig.~\ref{perfil} we plot the fractional abundances of the most
abundant ions of oxygen for a particular model of our grid, vs. the radial
thickness, which is the distance to the
inner face of the ionized shell. 
The continuum line represents O$^{0}$, the dashed line O$^{+}$,
and the dot-dashed line O$^{++}$. As shown in the figure, the O$^{++}$ ion
is dominant throughout most of the radial profile of the nebulae. It shows a
sharp cut-off  almost at the external boundary, where the O$^{+}$ and
especially O$^{0}$ become dominant. At the edge of the ionized nebula,
even the O$^{+}$ ion decays. Although the width of the O$^{0}$
region and the steepness of the transition strongly depend on the
physical properties of the nebula (ionization parameter, shape of the
stellar continuum, metallicity) the qualitative behaviour is essentially
as described above for all cases. The lower plot shows the emissivity of
the O$^{0}$ ion as a function of radial thickness. It can be seen that
the maximum is reached in the outermost ionized layers. So this
plot is telling us that O$^{0}$ is an excellent tracer of the H{\sc
ii} region boundary. The emergent flux of
the [O~{\sc i}] $\lambda$6300 \AA\ emission line should decrease as the escape of
Lyman continuum photons increases.

%************************************

%Aqui incluir alguna referencia sobre el efecto de los choques en la linea de
%[OI]

%***************************************

Figure~\ref{hiiboundhi} shows the [O~{\sc i}]/H$\beta$ ratio as a
function of [O~{\sc ii}]/H$\beta$ for our set of models. Given the
range in age measured for optically detected H~{\sc ii} regions (see
Martin \& Friedly 1999), only models
with ages of between 2 and 9~Myr are represented. Plot (a)
corresponds to models with metallicities equal to or greater than solar,
and plot (b) corresponds to models with subsolar abundances.
In both plots, solid, dashed and dot-dashed lines represent the
ionization-bounded models with different IMFs. Each line corresponds to the subset of
models with a given age, IMF, and metallicity, covering all possible
values of the ionization parameter.
Crosses and diamonds represent those density-bounded models with fractions of
escaping Lyman continuum photons of 20\% and 40\% respectively; these were
plotted in order to compare their position in the diagram  to  that of the
ionization-bounded models.

As can be seen from Fig.~\ref{hiiboundhi}(a), most ionization-bounded
high-metallicity models, follow a linear
trend in the plot with a slope of
 approximately unity. These models correspond to H~{\sc ii} regions for
which most of the oxygen is doubly ionized,  the O$^{++}$
zone almost reaching   the outer limit of the nebula. The role of the ionization parameter is limited in regulating
the relative widths of the O$^{+}$ and O$^{0}$ zones of the nebula,
which are confined to the outermost layers. 
Thus, as soon as the
degree of ionization decreases, the O$^{0}$ fraction increases producing a dramatic
decrease in the O$^{+}$ fraction, which is far less important than that of
O$^{++}$. Thus, the fluxes of the emission lines from both ions
are linearly related.

For other ionization-bounded models, the [O~{\sc i}]
and [O~{\sc ii}] fluxes are almost unrelated and correspond to H~{\sc
ii} regions where the O$^{+}$ ion is dominant throughout, 
except in the outer boundary, where the O$^{0}$ is
always  dominant. The contribution of the [O~{\sc ii}] line comes 
from a highly extended region compared to that of the [O~{\sc i}] line,
and thus the flux of the [O~{\sc ii}] line remains almost unchanged
with small changes in the flux of the [O~{\sc i}] line.

The two different trends shown by the
ionization-bounded models is understood in terms of 
the presence of an excess of energetic photons in the ionizing SED
coming from Wolf--Rayet (WR) stars. In fact, these are  models for which the
fraction of WR to O stars is non-negligible, and which show the linear
unity-slope trend previously mentioned. These stars appear only 
for a short time period of between about $3 \times 10^{6}$ to $6.5
\times 10^{6}$~yr and contribute significantly to the total number of 
ionizing stars for metallicities of the order of solar or
higher (see Leitherer \& Heckman 1995 for details). Of course, the
assumed upper mass limit of the IMF plays an important role in determining
 whether or not models
with the same physical parameters  follow the linear trend,
since for an IMF with $M_{\rm up} = 30\ M_{\odot}$ the beginning 
of the WR phase is delayed by some 3 Myr. However, for the range of
ages displayed in the plot, this result holds regardless of the upper
mass limit assumed for the IMF. 

The density-bounded models tend to show almost similar fluxes in the
[O~{\sc ii}] line to those corresponding to ionization-bounded models, but
with significantly lower fluxes in the [O~{\sc i}] line, and thus populate
the region below this linear trend. 
There is significant overlapping between both ionization-bounded and
density-bounded models.
However, there is an area
below the linear trend---the shaded region in the plot---with no
overlap between both sets of models and populated only by
density-bounded models. The existence of only density-bounded
models in this region suggests that H~{\sc ii} regions populating this
region should be the density-bounded ones.

The result is not so clear for low metallicity models. 
Figure~\ref{hiiboundhi}(b) shows that both sets of models
appear to be mixed throughout the diagram, and only a very small
region---shaded in the plot---appears to be populated only by
density-bounded models, thus making this diagnostic
useless.

Another process that can play a role in the intensity of the [O~{\sc
i}] $\lambda$6300 \AA\ line is the presence of shock fronts due to galactic
winds and/or supernovae explosions. However, this mechanism tends to increase
the intensity of the [O~{\sc i}] $\lambda$6300 \AA\ line (Dopita \& Sutherland
1996), thus producing the opposite effect to that produced by UV photon
escaping from the nebulae. This means that the escaping photons fractions
estimated from our models are lower limits to the real values if we assume the
likely presence of shocks in these regions.

\section{Discussion and Conclusions \label{discon}}

In order to check the validity of the diagram presented in the
previous section, we have used a large sample of observed H~{\sc ii} regions.
The sample selected (van Zee et al. 1998) comprises spectra for 186 H~{\sc
ii} regions belonging to 13 nearby spirals. In
addition, the [O~{\sc i}] $\lambda$6300 \AA\ line has been measured for many
regions. For each region, abundances values  are provided. 
We have selected those regions for which a value of $12 + \log
\rm{O/H} > 8.7$ has been reported. We thus include the H~{\sc ii} regions with
metallicity higher than or of the order of solar, taking into account that the
typical errors in the semi-empirical calibrations for the oxygen abundances are
of the order of 0.10~dex. 

In particular, the high-metallicity regions are located towards the centres of
spiral galaxies (see van Zee et al. 1998 and references therein for comments
about the metallicity gradients measured for their sample galaxies). Also,
studies of complete populations of H~{\sc ii} regions in spiral galaxies show
that most of them are located in the inner regions with a maximum at about
0.25º $R_{25}$ with an outward exponential decrease (Rozas et al. 1996). In
addition, Fergusson et al. (1998) showed that although H~{\sc ii} regions can be
found as far out as two optical radii, they appear small and faint compared to
their inner-disk counterparts. So taking all the three remarks together suggests
that our diagnostic is of general utility for H~{\sc ii} regions in spiral
galaxies, and that many of the brightest ones should fulfil the requirements to
be checked with our diagnostic. 
%In particular, the brightest density-bounded
%H{\sc ii} regions should account for most of the escaping Lyman continuum
%photons in the spiral galaxies.

Figure~\ref{todas} shows the [O~{\sc ii}] vs. [O~{\sc i}] diagnostic
diagram for the selected subsample of H~{\sc ii} regions. 
Thick crosses represent the observed
H~{\sc ii} regions. Error bars from van Zee et al. (1998) are also plotted.
H~{\sc ii} regions with no detected emission from the [O~{\sc
i}] $\lambda$6300 \AA\ are represented with downward-pointing arrows. 
Solid lines correspond
to the ionization-bounded models. The shaded region corresponds to
the area occupied by density-bounded models with no overlap with
ionization-bounded models.

As this figure shows, many observed H~{\sc ii} regions lie within the
shaded region which corresponds to the area occupied by density-bounded
models and with no overlapping with the ionization-density models (in
particular, 80\% of those for which the [O~{\sc i}] $\lambda$6300 \AA\ line was
detected). If we also take into account the regions for which this line was not
detected, the fraction of H~{\sc ii} regions lying within the shaded region
or just below amounts to 70\%.
According to the position occupied by the density-bounded models in Fig.~3(a), the
estimated fraction of  Lyman continuum photons escaping from the H~{\sc ii}
regions falling in the shaded region ranges between 20\% and 40\% although this
proportion
could be even more pronounced if other mechanisms that enhance the intensity of
the [O~{\sc i}] line are present.
Of course, a more precise determination of the fraction
of escaping Lyman continuum photons  needs an accurate fit of the
complete spectra of the observed H~{\sc ii} regions to individual models.

Concerning the negative detections, 64\% of these fall just below the shaded
regions, which means that they fulfil the applicability conditions of our
diagnostic on detection of the [O~{\sc i}] $\lambda$6300 \AA\ line. If the
absence of the [O~{\sc i}] $\lambda$6300 \AA\ line is caused by lack of sensitivity,
nothing can be said about the fraction of escaping UV photons; but if it is caused by real density-boundedness, fractions of UV photons even greater than 40\% could
be escaping from the H~{\sc ii} region. In any case, more detailed studies of 
individual spectra are required for these H~{\sc ii} regions.

We  should emphasize  that the available data for the H~{\sc ii} regions in the
sample that we have used are integrated and thus do not take into account
the possible irregular morphologies of the regions. Therefore, it could be that
H~{\sc ii} regions are only partially density bounded, probably on the side
orientated towards lower densities in the molecular clouds, and that what we are
measuring is just an average effect. In order to go deeper into the study  of
the structure and porosity of H~{\sc ii} regions, spectral data on the
individual shells of spatially resolved H~{\sc ii} regions are required in order
to check whether or not Lyman continuum photons are escaping through the ionized
shells.

In this paper we have presented evidence supporting 
the hypothesis that density-bounded H~{\sc ii} regions do
exist in spiral galaxies. The emission from the [O~{\sc
i}] $\lambda$6300 \AA\ line combined with that of the [O~{\sc
ii}] $\lambda\lambda$3727,3729 \AA\ doublet is a good diagnostic for discerning
 whether high metallicity H~{\sc ii} regions are density or
ionization bounded. A preliminary study of a sample of H~{\sc ii}
regions in spiral galaxies shows that some high metallicity H~{\sc ii} regions
confirm the predictions of density bounded photoionization models, 
with fractions of escaping Lyman continuum
photons ranging between 20\% and 40\% (but the fraction
could be higher for H~{\sc ii} regions
for which we do not have information about the intensity of the [O~{\sc
i}] $\lambda$6300 \AA\ line).
 However, it is not clear whether the
number of density-bounded regions and the fraction of Lyman continuum
photons escaping from them are enough to provide the observed ionization
of the diffuse medium, while we have no information on the
[O~{\sc i}] $\lambda$6300 \AA\ line for a complete sample of H~{\sc ii}
regions. 
Combined information on the spectra and H$\alpha$ luminosity of complete samples
of H~{\sc ii} regions in galaxies are needed to at least set a lower limit to the
number of ionizing photons escaping from H~{\sc ii} regions.
In fact, the breakout of ionized shells proposed by
Tenorio-Tagle et al. (1997), might be  an even more important contributor
to the ionization of the diffuse intergalactic medium.

\vspace{5mm}

%\acknowledgments

This study was partly financed by the Spanish DGI (AYA2001-3939-C03-03).
Thanks are given to G. Tenorio-Tagle, J. M. V\'{\i}lchez and J. Franco for
interesting comments and suggestions.
We also  acknowledge the Scientific Editorial Service of the IAC for 
corrections to versions of this article. JIP acknowledges the 5th
framework programme of human mobility of the UE for a Marie Curie fellowship.

\newpage

\begin{table}
\begin{center}
\begin{tabular}{lc}
\hline
Parameter & Values \\
\hline
$12 + \log \rm{O/H}$ & 8.3, 8.6, 8.9, 9.3 \\
$\log u$ & $-2.0$, $-2.5$, $-3.0$, $-3.5$, $-4.0$\\
$\alpha$ & $-2.35$, $-3.30$ \\ 
Age (Myr) & 1, 2, 3, 4, 5, 6, 7, 8, 9, 10\\
$M_{\rm up}~(M_{\odot})$ & 30, 100 \\
\hline
\end{tabular}
\caption{Physical parameters of the models.\label{models}}
\end{center}
\end{table}

\newpage

\clearpage

\begin{figure}
\begin{center}
\caption{The shaded area corresponds to the region occupied by the
measured H~{\sc ii} regions from McCall et al. (1985). Open squares correspond to
 models of our grid representing ionization-bounded H~{\sc ii}
regions. Filled squares correspond to the models representing
density-bounded H~{\sc ii} regions with the same physical
parameters.\label{o2o3}} \end{center}
\end{figure}

\newpage

\begin{figure}
\begin{center}
\caption{The upper plot shows the fractional abundances of the different
ions of oxygen as a function of the radial thickness for a particular
model of our grid. The lower plot shows the emissivity of the O$^{0}$ ion as a function
of the radial thickness for this model.\label{perfil}}
\end{center}
\end{figure}

\newpage

\begin{figure}
\begin{center}
\caption{(a) [O~{\sc i}]/H$\beta$ vs. [O~{\sc ii}]/H$\beta$ diagram for
ionization-bounded models (solid lines for $\alpha = -2.35$ and $M_{\rm up} =
100~M_{\odot}$, dashed lines for $\alpha = -2.35$ and $M_{\rm up} = 30~M_{\odot}$
and dash-dotted lines for $\alpha = -3.30$ and $M_{\rm up} = 100~M_{\odot}$). 
Crosses and diamonds represent density-bounded models with 20\% and 40\% 
escaping Lyman continuum photons
respectively. The shaded region represents the loci expected for the
density-bounded H~{\sc ii} regions. Only models with metallic abundances equal to or
greater than solar were considered in this plot.\label{hiiboundhi}}
\end{center}
\end{figure}

\newpage

\addtocounter{figure}{-1}

\begin{figure}
\begin{center}
\caption{(b) Same as Figure~\ref{hiiboundhi}(a) for models with
subsolar metal abundances.\label{hiiboundlo}} 
\end{center}
\end{figure}

\newpage

\begin{figure}
\begin{center}
\caption{Same  as Figure~\ref{hiiboundhi}. Solid lines
correspond to ionization-bounded models with metal abundances
equal to or greater than solar. Crosses correspond to the high-metallicity
H~{\sc ii} regions from van Zee et al. (1998). Downward-pointing arrows
represent those H~{\sc ii} regions with no flux reported for the [O~ {\sc
i}] $\lambda$6300 \AA\ line. The shaded region corresponds to the loci
occupied by density-bounded models.\label{todas}} 
\end{center}
\end{figure}

\end{document}